\documentclass[sigplan,10pt,screen]{acmart}

\usepackage[normalem]{ulem}
\usepackage{listings}
\usepackage{xparse}
\usepackage{braket}
\usepackage{qcircuit}
\usepackage{stmaryrd}
\usepackage{mathtools}
\usepackage{bussproofs}
\usepackage[capitalise]{cleveref}
\usepackage{comment}
\usepackage{amsmath}
\usepackage{graphicx}
\usepackage{amsfonts}  
\usepackage{stmaryrd}
\usepackage{arydshln}
\usepackage{appendix}
\usepackage[framemethod=tikz]{mdframed}
\usepackage{parcolumns}
\usepackage[scaled]{beramono}
\usepackage{sidecap}
\usepackage{letltxmacro}
\usepackage[bottom]{footmisc}
\usepackage{mathpartir}
\usepackage{wrapfig}
\usepackage{subcaption}

\usepackage[font=small,labelfont=bf]{caption}

\setcopyright{acmlicensed}
\acmPrice{15.00}
\acmDOI{10.1145/3519939.3523431}
\acmYear{2022}
\copyrightyear{2022}
\acmSubmissionID{pldi22main-p34-p}
\acmISBN{978-1-4503-9265-5/22/06}
\acmConference[PLDI '22]{Proceedings of the 43rd ACM SIGPLAN International Conference on Programming Language Design and Implementation}{June 13--17, 2022}{San Diego, CA, USA}
\acmBooktitle{Proceedings of the 43rd ACM SIGPLAN International Conference on Programming Language Design and Implementation (PLDI '22), June 13--17, 2022, San Diego, CA, USA}


\definecolor{dark-gray}{gray}{0.2}
\definecolor{mygreen}{HTML}{467E7E}
\definecolor{mygray}{rgb}{0.5,0.5,0.5}
\definecolor{mymauve}{HTML}{AC134D}
\definecolor{myblue}{HTML}{4169E1}
\definecolor{myorange}{HTML}{B25A00}

\newenvironment{revise}{}{\ignorespacesafterend}
\newenvironment{reviseg}{}{\ignorespacesafterend}

\newcommand{\framework}{Giallar}
\newcommand{\qiskit}{Qiskit}
\newcommand{\numPassVerified}{44}
\newcommand{\totalPassVerified}{56}
\LetLtxMacro\oldttfamily\ttfamily
\DeclareRobustCommand{\ttfamily}{\oldttfamily\csname ttsize\endcsname}
\newcommand{\setttsize}[1]{\def\ttsize{#1}}%
\setttsize{\small}

\newcommand{\qceq}{\push{\rule{.3em}{0em}\equiv\rule{.3em}{0em}}}

\NewDocumentCommand{\codeword}{v}{%
\texttt{\textbf{\textcolor{blue}{#1}}}%
}

\definecolor{mygray}{rgb}{0.5,0.5,0.5}

\lstdefinestyle{qubit}{
  language=Python,
  backgroundcolor=\color{white},   %
  basicstyle=\linespread{0.9}\ttfamily\footnotesize,        %
  breakatwhitespace=false,         %
  breaklines=true,                 %
  commentstyle=\color{mygreen},    %
  deletekeywords={...},            %
  escapeinside={<@}{@>},          %
  extendedchars=true,              %
  keepspaces=true,                 %
  keywordstyle=\bfseries\color{myblue!70},       %
  language=Octave,                 %
  otherkeywords={uint, RELATIONAL_SEMANTICS, float, void,@classmethod, CLIGHTX_SEMANTICS,fresh\_int,module,class,...,import,from,def, Map, Mapper, DAGCircuit ,None, assertion , terra\_call},     
  deletekeywords={get,angle, gamma, invariant,input,size,delete},
  emph = { certiq_prove, match, end, with, let,ret, do, forall, exists, not, N,const,  list, Fixpoint, Function, Definition,Goal, Lemma, CNOT,H,Some, Rz, ForAll, And, Measure,Rx,Meas, equiv, Bloch_rep, CouplingMap, Layout, __init__, True, False, Implies, prove, BasicSwap, count_call, assert, assume,input,output,next_gate},
  emphstyle=\bfseries\color{myorange!70},%
  showspaces=false,                %
  showstringspaces=false,          %
  showtabs=false,                  %
  stringstyle=\color{mymauve},     %
  tabsize=1,	    
 morecomment=[f][\bfseries\color{mymauve!70}][0]{@},
 morecomment=[f][\itshape\color{mygreen}][0]{//},
  numbersep=2pt,                   %
  numberstyle=\tiny\color{mygray}, %
}

\newcommand{\para}[1]{\vspace{5pt}\noindent\textbf{\textit{#1.\ }}}

\begin{document}

\title{\framework{}: Push-Button Verification for the Qiskit Quantum Compiler}
\author{Runzhou Tao}
\affiliation{
  \institution{Columbia University}            %
  \city{New York}
  \state{NY}
  \country{USA}                    %
}
\email{runzhou.tao@columbia.edu}          %

\author{Yunong Shi}
\affiliation{
  \institution{Amazon}            %
  \city{New York}
  \state{NY}
  \country{USA}                    %
}
\email{shiyunon@amazon.com}          %

\author{Jianan Yao}
\affiliation{
  \institution{Columbia University}            %
  \city{New York}
  \state{NY}
  \country{USA}                    %
}
\email{jianan@cs.columbia.edu}          %
\author{Xupeng Li}
\affiliation{
  \institution{Columbia University}            %
  \city{New York}
  \state{NY}
  \country{USA}                    %
}
\email{xupeng.li@columbia.edu}          %

\author{Ali Javadi-Abhari}
\affiliation{
  \institution{IBM Research}            %
  \city{Yorktown Heights}
  \state{NY}
  \country{USA}                    %
}
\email{ali.javadi@ibm.com}          %

\author{Andrew W. Cross}
\affiliation{
  \institution{IBM Research}            %
  \city{Yorktown Heights}
  \state{NY}
  \country{USA}                    %
}
\email{awcross@us.ibm.com}          %

\author{Frederic T. Chong}
\authornote{Frederic T. Chong  is also Chief Scientist at Super.tech and an advisor to Quantum Circuits, Inc.}
\affiliation{
  \institution{The University of Chicago \and Super.tech}
  \city{Chicago}
  \state{IL}
  \country{USA}                    %
}
\email{chong@cs.uchicago.edu}          %

\author{Ronghui Gu}
\authornote{Ronghui Gu is also the Founder of CertiK.}
\affiliation{
  \institution{Columbia University \and CertiK}            %
  \city{New York}
  \state{NY}
  \country{USA}                    %
}
\email{ronghui.gu@columbia.edu}          %

\renewcommand{\shortauthors}{R. Tao, Y. Shi, J. Yao, X. Li, A. Javadi-Abhari, A. Cross, F. Chong, and R. Gu}

\begin{abstract}

This paper presents \framework{}, a fully-automated verification toolkit for quantum compilers. \framework{} requires no manual specifications, invariants, or proofs, and can automatically verify that a compiler pass preserves the semantics of quantum circuits.
To deal with unbounded loops in quantum compilers, \framework{} abstracts three loop templates, whose loop invariants can be automatically inferred. 
To efficiently check the equivalence of arbitrary input and output circuits
that have complicated matrix semantics representation, \framework{} introduces a symbolic representation for quantum circuits and a set of {\reviseg rewrite} rules for
\begin{reviseg}
showing the equivalence of 
\end{reviseg}
symbolic quantum circuits. 
With \framework{}, we implemented and verified \numPassVerified{} (out of 56) compiler passes in 13 versions of the \qiskit{} compiler, the open-source quantum compiler standard, during which three bugs were detected in and confirmed by \qiskit{}.
Our evaluation shows that  
most of  \qiskit{} compiler passes can be automatically verified in seconds
and  verification imposes only  a modest overhead to  compilation performance. %
 \end{abstract}
\maketitle

\section{Introduction}
\label{sec:introduction}

Quantum compilers are  essential components in the quantum software stack, bridging quantum applications to hardware. 
However, correctly implementing a quantum compiler is difficult due to 
two reasons: 1) quantum compilers have to obey quantum mechanics rules that are highly nonintuitive~\cite{zam2016talk};
and 2) quantum compilers must perform heavy optimizations
to fit quantum programs into real quantum devices
of limited qubit lifetime and connectivity~\cite{sabre}. 
These challenges make quantum compilers error-prone. For example, Bugs4Q~\cite{bugs4q} finds 27 bugs in the \qiskit{} compiler~\cite{terra_issue}, the most widely-used open-source quantum compiler with more than 100K users from 171 countries.
Undetected bugs in the \qiskit{} compiler can corrupt 
the computation results of the millions of 
simulations and real runs on quantum devices.
Eliminating bugs in quantum compilers becomes  crucial  for the success of near-term quantum computation.

In this paper, we present \framework{}, a toolkit that helps programmers write quantum compiler passes
and formally verify their correctness in a \emph{push-button} manner.
\framework{} requires \emph{no} 
manual annotations, invariants, specifications, or proofs about
the implementation of the quantum compiler pass.
\framework{} performs verification to check if the compiler pass preserves
the semantics of quantum circuits.
An executable pass implementation is produced if the verification succeeds.
If there is a bug, \framework{} produces
a counterexample to help identify and fix the cause.

The main challenge in
applying formal verification to
building correct quantum compilers
is to minimize proof burden.
Recent efforts~\cite{voqc, Amy2017} 
have shown that it is feasible to manually verify the correctness of
quantum compiler 
passes using  interactive theorem provers such as Coq~\cite{Coq12}.
However, writing such proofs requires
a significant time investment from formal verification experts,
and the size of proofs 
can be several times or even more than an order of magnitude
larger than that of {\reviseg the compiler} implementation,
making the proofs expensive to develop and maintain.
For example, \citet{voqc} reported that verifying 
circuit mapping, a single transformation pass consisting of 70 lines of code,
requires writing 2,100 lines of proof{\reviseg s}.
These verification frameworks are impractical
for verifying fast-moving and frequently changing 
quantum compilers such as \qiskit{}, 
which has 683 commits on the main branch from 64 contributors in the first 10 months of {\reviseg the year} 2021~\cite{qiskit_repo}.

To allow non-formal-verification-experts to develop  correct
quantum compilers without such a proof burden,
\framework{} provides fully automated reasoning.
Conceptually, showing that a quantum compiler pass is correct 
involves proving that it preserves the semantics for any input quantum circuit
in all possible execution paths.
In practice, automating such a proof 
 faces  \emph{classical} and \emph{quantum} challenges.

On the one hand, a quantum compiler
is  a classical program 
{\reviseg that} intensely uses unbounded loops
and {\reviseg complex utility functions (containing nested loops and recursions)}  
to perform transformations and 
optimizations based on the information of the whole circuit.
These program features in general
are hard to reason about automatically.
For example, automated verification frameworks such as Alive~\cite{alive} and Hyperkernel~\cite{hyperkernel} require the input program to be loop-free and recursion-free, or only have bounded loops. 

On the other hand,
the correctness of a quantum compiler pass is usually defined as
the semantics preservation property for quantum circuits,
while efficient equivalence checking for general quantum circuits 
is still beyond reach.
Previous quantum verification work{\reviseg s}~\cite{voqc,qbricks,Amy2019}  %
{\reviseg rely} on users {\reviseg manually} reasoning about the equivalence
of quantum circuits either using the denotational semantics~\cite{NielsenChuang} (i.e., the matrix representation), which requires exponential time and memory to compute,
or  using the path-sum semantics~\cite{Amy2019}, which only supports a restricted subset of quantum states. 
Neither of these approaches is feasible for automated verification of quantum compilers.

Our \framework{} toolkit
addresses the {\reviseg above} challenges  
by leveraging domain-specific knowledge of quantum compilation.
First, most of the unbounded loops in quantum compilers follow one of a few specific patterns
to traverse the input quantum circuit. %
\framework{} 
abstracts these patterns 
into three \emph{loop templates} for users to write unbounded loops,
whose loop invariants can be automatically inferred without any user input.
The compiler pass 
containing unbounded loops 
then becomes symbolically executable
by {\reviseg reducing} the loops with the inferred  invariants.
For each loop, \framework{} will also generate a separate proof goal that the symbolic execution
of the 
loop body indeed retains the inferred invariants.
\framework{} formulates  {\reviseg such proof goals} as  SMT problems and invokes Z3~\cite{z3} to solve them.
Our three loop templates  cover all the unbounded loops
in all verified %
\qiskit{} passes,
while new loop templates can be easily introduced to meet future needs.
\begin{reviseg}
As for complex utility functions that contain nested loops and recursions,
\framework{} provides a verified library of utility functions
that is shared by multiple passes,
such that  their invocations
can be replaced by their specifications
during the symbolic execution.
\end{reviseg}

Second, instead of directly using the matrix representation
to check the equivalence of the input and output quantum circuits,
\begin{reviseg}
\framework{} shows that the output quantum circuit can be obtained from
the input circuit through a sequence of \emph{equivalent rewrites}.
To define such {\reviseg rewrite rules},
{\revise we define symbolic representation and execution} 
for quantum circuits, which are different from
the symbolic execution of the compiler implementation mentioned above.
All {\reviseg rewrite} rules operate the symbolic quantum circuit
and preserve the results of the symbolic execution.
The set of {\reviseg rewrite} rules provided by \framework{} is general enough to cover common quantum compilations and small enough to enable efficient checks.
The {\reviseg rewrite} rules
are manually verified
in the Coq proof assistant~\cite{Coq12},
and the verification is done once and for all.
\end{reviseg}

Third, in contrast to existing automated verification frameworks~\cite{alive,yggdrasil, serval}
that require users to provide specifications,
\framework{} introduces a set of Python virtual classes for different types of passes in \qiskit{}.
\framework{} can automatically generate proof obligations for the pass implementations inheriting these virtual classes.

We have used \framework{} to implement and verify \numPassVerified{} out of 
\totalPassVerified{} compiler passes in 13 versions (from v0.19 to v0.32) of the \qiskit{} compiler.
Among  12 failed passes, 
eight passes deal with pulse-level behaviors,
two passes rely on external solvers,
one pass involves a randomized routing algorithm,
and one pass produces an approximated circuit within a given error bound. 
These passes are not supported by \framework{},
and are also costly or infeasible to manually verify
using existing quantum verification frameworks~\cite{voqc,qbricks,Amy2017,Amy2019}.
During the verification,
we found three critical bugs in (and confirmed by) {\reviseg the \qiskit{} team}, two of which are unique to quantum computing. 
Our evaluation shows that most of the \qiskit{} compiler passes can be automatically verified in seconds
and the verified compiler passes only have modest performance overhead compared with the unverified \qiskit{} implementation. 

This paper makes the following contributions: 
\begin{itemize}
 \setlength\itemsep{1em}
    \item The \framework{} toolkit that allows non-formal-verification-experts to 
    build provably correct quantum compilers
    in the face of frequent
    changes and new features.
    
    \item %
        A domain-specific approach using loop templates, virtual classes for passes,
        and verified utility library to fully automate the verification of 
        quantum compilers.
        
    \item A set of verified {\reviseg rewrite} rules that enables the efficient equivalence checking 
    for quantum circuits.

    \item A case study of implementing and verifying the compiler passes  of \qiskit{},
    the most widely-used quantum compiler, using \framework{}. 
    Three critical bugs have been found during the verification and confirmed by \qiskit{}.
\end{itemize}

\section{Background}
\label{sec:background}
We first introduce the necessary background on the verification of quantum computing and quantum compilation. For more details of quantum computing, please refer to \cite{r3}.

\subsection{Quantum Basics}
\label{sec:quantumbasics}
Quantum states are represented as 2-dimensional complex vectors, named qubits, e.g., $\ket{0} = (1,0)^T$ and $\ket{1} = (0,1)^T$. Quantum gates are operations of quantum states that can be represented by unitary matrices (see {\reviseg Figure~\ref{fig:gatematrix}}). In contrast with classical computing, an n-qubit state (or gate) is represented by a $2^n$-dimensional vector (or a $2^n \times 2^n$ vector), leading to the exponential cost in space and time required to directly simulate quantum programs using matrix and vector representations.

\begin{figure}[t]
        \small

        \vspace{3pt}
        \begin{tabular}{ccc}
        \centering
           
                \begin{minipage}{0.10 \textwidth}
                \centering 
            \Qcircuit @C=0.3em @R=0.2em{
                 &\gate{X}& \qw \\
            }
            \end{minipage}
           &
            
            \begin{minipage}{0.10 \textwidth}
                \centering
            \Qcircuit @C=0.3em @R=0.2em{
                 &\gate{H}& \qw \\
            }
            \end{minipage}&

\begin{minipage}{0.10 \textwidth}
    \centering
    \Qcircuit @C=0.3em @R=0.2em{
     &\ctrl{3}  &\qw  \\
&&  \\
\\
         &\targ     &\qw  \\
}
\end{minipage}\vspace{10pt} \\

 $ \llbracket X \rrbracket = \begin{bmatrix} 0 & 1 \\ 1 & 0 \end{bmatrix}$ & $ \llbracket H \rrbracket =\frac{1}{\sqrt{2}}\begin{bmatrix} 1 & 1 \\ 1 & -1 \end{bmatrix}$ & $ \llbracket CX \rrbracket= \begin{bmatrix} 1 & 0 & 0 & 0 \\ 0 & 1 & 0 & 0 \\ 0 & 0 & 0 & 1 \\ 0 & 0 & 1 & 0 \end{bmatrix}$ 
        \end{tabular}
        \caption{ Circuit diagram symbols (top) and the denotational semantics (bottom) for several 1-qubit and 2-qubit gates.}
        
        \label{fig:gatematrix}
\end{figure}

\subsection{Quantum Program}
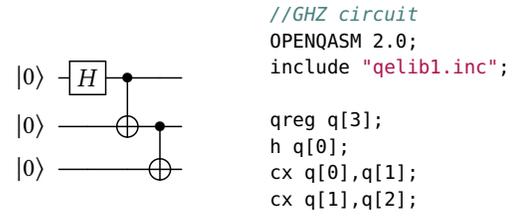
\begin{figure}[t]

 \begin{minipage}{0.18\textwidth}
\Qcircuit @C=0.41em @R=0.8em {
&   \lstick{\ket{0}} &\gate{H}  &\ctrl{1}   &\qw        &\qw\\
&\lstick{\ket{0}} &\qw       &\targ      &\ctrl{1}   &\qw\\
&   \lstick{\ket{0}} &\qw       &\qw        &\targ      &\qw\\
}
\end{minipage} \hspace*{1cm}
\begin{minipage}{0.18\textwidth}
\centering
\begin{small}
    \begin{lstlisting}[style=qubit]
//GHZ circuit
OPENQASM 2.0;
include "qelib1.inc";

qreg q[3];
h q[0];
cx q[0],q[1];
cx q[1],q[2];
    \end{lstlisting}
\end{small}
\end{minipage}\hspace*{0.2cm}
\vspace{-10pt}
\caption{Circuit diagram (left) and the OPENQASM IR (right) of a simple GHZ circuit \cite{ghz}.
}
\label{fig:qasm}
\end{figure}

Quantum programs %
process quantum states with quantum gates.
They are often represented graphically with gates as nodes and qubits as wires (see {\reviseg Figure~\ref{fig:gatematrix}}).
The most widely-used quantum programming language is OpenQASM~\cite{cross2017open}. Figure~\ref{fig:qasm} shows an example of a 3-qubit circuit in graphical representation and in OpenQASM.

Programs considered in \framework{} follow a variant of the OpenQASM language as below.
\begin{align*}
     P &:= \texttt{skip} \\
     &\mid U(q_1, \ldots, q_n) \\
     &\mid P_1 ; P_2
\end{align*}
Empty circuit is denoted as \texttt{skip}; applying an n-qubit gate on selected qubits $q_1,...q_n$ is denoted as $U(q_1, \ldots, q_n)$; concatenation of two circuits $P_1$ and $P_2$ is denoted as $P_1 ; P_2$.
Features in OpenQASM that are not supported by existing hardware such as classical control flow are not included in our syntax. Nevertheless, this syntax is general enough to support a wide range of gate representations, circuit transformations, and various targeting hardware. This restriction on syntax is also a common practice in previous work on manual verification of %
quantum compilers~\cite{voqc}.
The input and output of each quantum compilation pass are both quantum programs. %

\begin{figure}[t]
    \centering\small
    \begin{align*}
    \llbracket \texttt{skip} \rrbracket_{\textit{nqreg}}:=&\quad I_{\textit{qreg}} \\
     \llbracket U\rrbracket_{\textit{nqreg}}:=&\quad\texttt{matrix}(U_{q_1,...,q_n}) \otimes I_{\textit{qreg} \backslash \{q_1, \ldots, q_n\}} \\
        \llbracket C_1\ ; C_2\rrbracket_{\textit{nqreg}}:=&\quad\llbracket C_1\rrbracket_{\textit{nqreg}} \times \llbracket C_2\rrbracket_{\textit{nqreg}}
    \end{align*}
    \vspace{-10pt}
    \caption{Denotational semantics of quantum circuits in \framework{}, %
    where \texttt{matrix} denotes the unitary matrices of the quantum operations
    and \texttt{qreg} 
    denotes the set of qubit
    registers.
    }
    \label{fig:denotational}
\end{figure}
    
\para{Denotational semantics} 
Figure~\ref{fig:denotational} shows
the denotational semantics of a quantum circuit $C$,
which is defined as its corresponding unitary matrix
and is denoted as $\llbracket C \rrbracket_{\textit{nqreg}}$,
where ${\textit{nqreg}}$ is the number of qubits in the quantum register used in the circuit.
The denotational semantics of an empty circuit 
with ${\textit{nqreg}}$ qubits is the identity matrix of size ${\textit{nqreg}}$. The semantics for a quantum gate is the tensor product of its matrix representation on its qubit operands and the identity matrix on other unrelated qubits. 
The semantics of the concatenation of two quantum programs is the multiplication of their matrix representations.

\begin{figure}[t]
    \centering
    \begin{subfigure}[b]{0.20\textwidth}
    \centering $$
    \Qcircuit @C=0.8em @R=0.7em {
&\ctrl{2}   &\qw        &\targ        &\qw\\
&\qw        &\ctrl{1}   &\ctrl{-1}    &\qw\\
&\targ      &\targ      &\gate{R_z}   &\qw
}$$
\caption{The original circuit}  
    \end{subfigure}
    \begin{subfigure}[b]{0.20\textwidth}
    \centering
\begin{tikzpicture}
\node[circle, draw] (a) at (0, 0) {\color{blue}{$a$}};
\node[circle, draw] (b) at (1, 0) {\color{red}$b$};
\node[circle, draw] (c) at (2, 0) {\color{purple}$c$};
\draw (a) -- (b);
\draw (b) -- (c);
\end{tikzpicture}
\caption{Physical qubits}
    \end{subfigure}\\
    \begin{subfigure}[b]{0.20\textwidth}
    \centering$$
    \Qcircuit @C=0.8em @R=0.7em {
\lstick{\color{blue}a} &\ctrl{2}   &\qw        &\targ         &\qw\\
\lstick{\color{red}b}&\qw        &\ctrl{1}   &\ctrl{-1}       &\qw\\
\lstick{\color{purple}c}&\targ      &\targ      &\gate{R_z}   &\qw
}$$
\caption{After layout selection}  
    \end{subfigure}
    \begin{subfigure}[b]{0.20\textwidth}
    \centering$$
    \Qcircuit @C=0.8em @R=0.7em {
&\qw            &\ctrl{1}   &\qw        &\qw
&\qw        &\targ  &\qw\\
&\qswap         &\targ      &\targ      &\gate{R_z}
&\qswap      &\ctrl{-1}      &\qw\\
&\qswap\qwx[-1] &\qw        &\ctrl{-1}  &\qw
&\qswap\qwx[-1] &\qw        &\qw
}$$
\caption{After routing}
    \end{subfigure}
    \vspace*{0.15cm}
    \caption{An example of layout selection and routing passes acting on a quantum circuit. (a) The original quantum circuit. Each line represents a logical qubit. (b) The target quantum computer's qubit arrangement, lines denote that two-qubit gates are allowed between the qubit pair. (c) After layout selection, the logical qubits in the circuit get mapped to physical qubits. At this point, the circuit is still not 
   runable on the target quantum computer because there is a 2-qubit gate between a and c. (d) After routing, swap gates are added into the circuit so that all 2-qubit gates are allowed. }
    \label{fig:translation}
\end{figure}
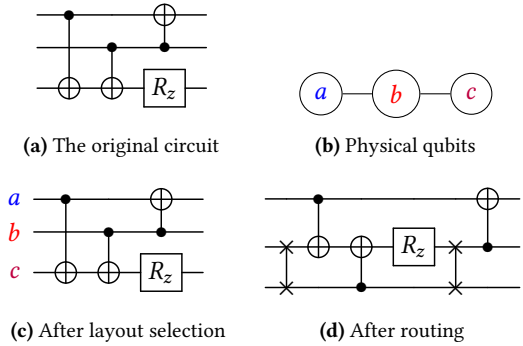
\subsection{Quantum Compilation}
\label{sec:qcompilation}
Quantum compilation is the process of translating 
high-level quantum circuit descriptions 
into optimized low-level circuits that are executable on  hardware. 
Most quantum compilers follow a design philosophy resembling that of LLVM~\cite{Lattner}:  circuit IRs are sequentially fed into a cascade of compiler components, called \emph{passes}, to be transformed and optimized.
The \qiskit{} compiler
has \emph{seven} types of compiler passes: layout selection, routing, basis change, optimizations, circuit analysis, synthesis, and additional assorted passes. 

\para{Layout selection and routing passes}
A layout selection pass maps logical qubits in the program to physical qubits on a specific hardware. A routing pass ensures that the quantum circuit conforms to the topological constraints of the quantum hardware. For example, in {\reviseg Figure~\ref{fig:translation}}, the quantum hardware has 3 physical qubits $a$, $b$, and $c$ with the topology constraints that 2-qubit gates can only be preformed between $a-b$ and $b-c$. The layout selection pass first assigns the three logical qubits in the circuit to $a$, $b$, and $c$. The routing pass then inserts  swap gates so that all 2-qubit gates satisfy the topological constraints. Because CNOT gates cannot be preformed between $a$ and $c$, a swap gate between $b$ and $c$ is inserted before the first CNOT gate. To perform the last CNOT gate, $b$ and $c$ need to be swapped back. 

\para{Other passes}
A basis change pass helps with the decomposition of quantum circuits into the gate set supported by a target hardware backend. An optimization pass includes various circuit-rewriting-based optimizations such as gate cancellation~\cite{Maslov2008}, scheduling optimization~\cite{Shi2019}, noise adaptation~\cite{Murali2019}, and crosstalk mitigation~\cite{murali2020}. A circuit analysis pass does not modify the circuits but returns important information about the circuits. A synthesis pass performs large unitary matrix decomposition. Additional assorted passes perform miscellaneous tasks such as circuit validation. 

\subsection{The Z3Py Tool}
Z3Py supports the symbolic execution of sequential Python code without
loops and branches, and 
introduces two primitives \texttt{assume(cond)} and \texttt{assert(cond)}. The \texttt{assume(cond)} primitive adds
the condition \texttt{cond} into the assumption list. 
The \texttt{assert(cond)} primitive
calls the Z3 SMT solver to check whether the current symbolic execution state satisfies \texttt{cond} given the assumption list. For example, given the code snippet below,
\begin{center}
\begin{tabular}{@{}c|c}
\begin{lstlisting}[style=qubit]
assume(x >= 3)
y = x * x
assert(y > x)
z = y + 1
assert(z > 10)
\end{lstlisting} & 
\begin{lstlisting}[style=qubit]
x >= 3 added to assumptions
Symbolic execution: y = x * x
SMT check x >= 3 -> x * x > x success
Symbolic execution: z = x * x + 1
SMT check x >= 3 -> x * x + 1 > 10 fail
\end{lstlisting}
\end{tabular}
\end{center}

\noindent  Z3Py will output ``verified'' for the first assertion and a counterexample ``$x=3$'' for the second assertion.
\section{The \framework{} Workflow}
\label{sec:framework}

\begin{figure}[t]
    \centering
\begin{lstlisting}[style=qubit,numbers=left,xleftmargin=2em,framexleftmargin=1.5em]
import giallar
class SimpleCXCancellation(GeneralPass):
    def run(self, input):
        remain = input.copy()
        output = QCircuit()
        while remain.size() != 0:
            gate = remain[0]
            if gate.isCXGate():
                next = next_gate(remain, 0)
                g = remain[next]
                if g.isCXGate() and 
                   g.qubits == gate.qubits:
                    remain.delete(next)
                else:
                    output.append(gate)
            else:
                output.append(gate)
            remain.delete(0)
        return output
\end{lstlisting}
    \caption{A simplified implementation of the CXCancellation  pass. }
    \label{fig;basicswap_code}
\end{figure}

\begin{figure*}[t] 
    \centering
     \includegraphics[width=0.95\textwidth]{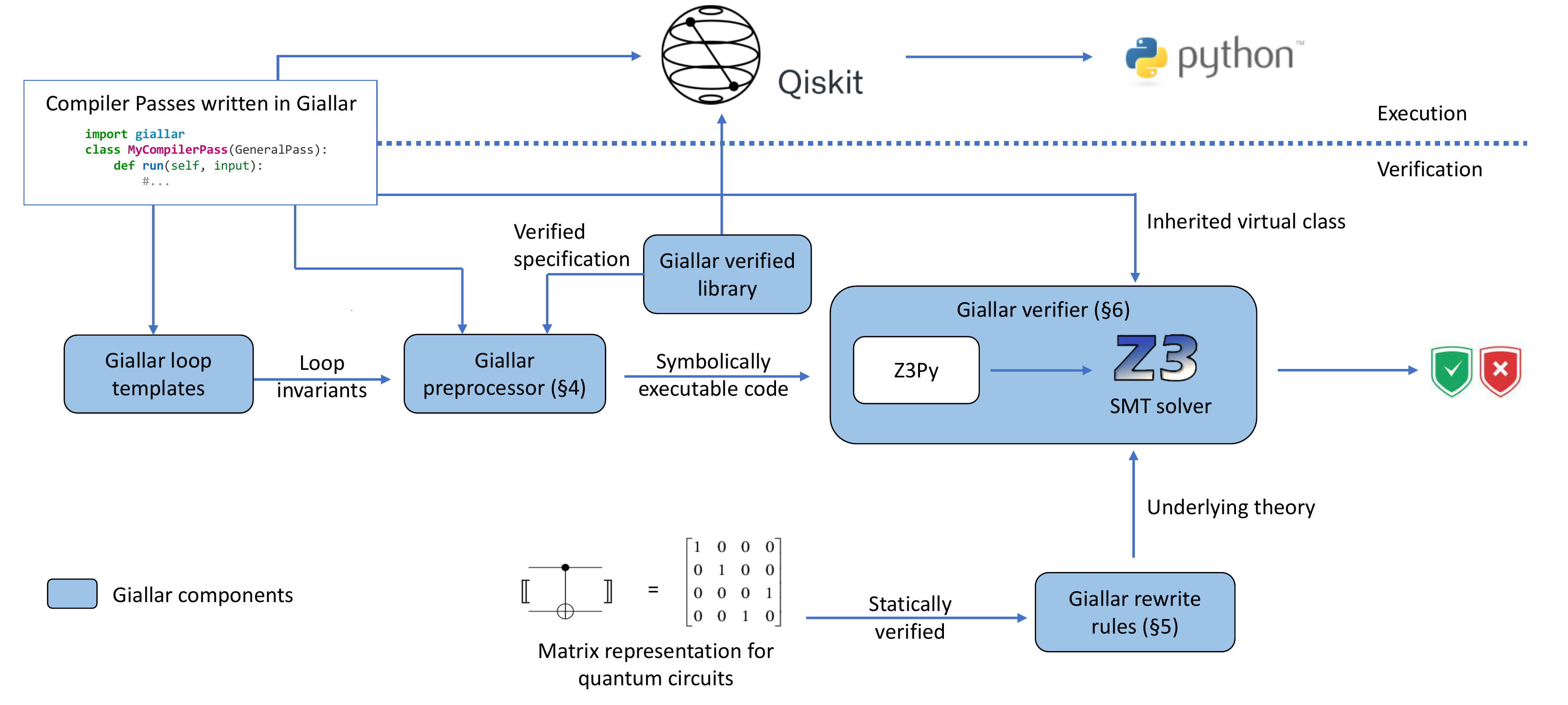}
    \vspace{-5pt}
    \caption{\framework{} workflow. %
    }
    \label{fig:\framework{}_flow}
\end{figure*}

The goal of the \framework{} toolkit is to allow quantum programmers
without much formal verification background
to write provably correct quantum compiler passes,
which can be readily integrated into the open-source \qiskit{} compiler and used in real-world quantum experiments. 

\framework{} consists of five major components
(see {\reviseg Figure~\ref{fig:\framework{}_flow}}): 
1) The \framework{} preprocessor that  {\reviseg translates} the compiler implementation into symbolically executable code;
2) A set of loop templates that can be used to 
infer loop invariants for 
unbounded loops in 
the pass implementation;
3) A verified library of %
utility functions;
4) A set of {\reviseg rewrite} rules for quantum circuits, {\reviseg enabling}
the efficient equivalence checking for quantum circuits;
5) The \framework{} verifier {\reviseg that}
generates the proof goal based on the type of the compiler passes
and verifies if the symbolic execution results satisfy the goal
using {\reviseg rewrite} rules.

In the rest of this section,
we will use a simplified version of  CXCancellation  pass
implemented in \framework{} (see {\reviseg Figure~\ref{fig;basicswap_code}})
as a running example to give a high-level overview
on \framework{}.
The CXCancellation pass is an optimization pass that 
scans through the gates of the input circuit
and cancels out adjacent CNOT (also called CX) gates that are on the same pair of qubits.
The variable \texttt{remain} represents the part of the circuit that 
has not been scanned yet.
During the scanning, if the current \texttt{gate} is not a CNOT gate, it will be directly appended to the \texttt{output} circuit. If the current \texttt{gate} is a CNOT gate,  the pass will then find the next gate in the remaining circuit that shares the qubits with the current \texttt{gate}, denoted as \texttt{g}. If \texttt{g} is also a CNOT gate on the same pair of qubits,  \texttt{gate} and \texttt{g} will be cancelled out and will not appear in the \texttt{output} circuit.

To enable the push-button verification for this pass,
we need to address two challenges:
1) the pass implementation contains an unbounded loop that forbids the automated symbolic execution,
and 2) checking the equivalence of input and output circuits
using denotational semantics requires 
to solve constraints over matrices with  arbitrary sizes,
which is infeasible for existing solvers.

Figure~\ref{fig:\framework{}_flow}
shows the workflow of the push-button verification for
the CXCancellation pass using \framework{}.
The key steps are described below.

\para{Infer loop invariants} 
The key to automate the symbolic execution
for unbounded loops is to infer the loop invariants,
which is hard for general-purpose programs
but is tractable for quantum compiler passes
with domain-specific knowledge.
For example, the unbounded loop in the 
the CXCancellation pass scans through the input circuit
to build the output circuit while maintaining a list of remaining
gates.
This loop pattern is quite common 
in many quantum compiler passes such as optimization passes.
For this loop pattern,
the loop invariant should be that the concatenation of 
the currently built part of the output circuit and the remaining gates must
be equivalent to the input circuit, i.e.,
their denotational semantics defined using
the matrix representations
are equivalent for any \textit{nqreg}:
$$
    [\![\texttt{?output};\texttt{?remaining\_gates}]\!]_\textit{nqreg} \equiv [\![\texttt{?input}]\!]_\textit{nqreg}.
$$

The \framework{} preprocessor 
will 
statically analyze the loop implementation and 
infer the assignment to the variable placeholders; for example,
the \texttt{remain} list should be assigned
to the variable placeholder ``\texttt{?remaining\_gates}.''
After the variable assignment, the invariant becomes:
$$
    [\![\texttt{output};\texttt{remain}]\!]_\textit{nqreg} \equiv [\![\texttt{input}]\!]_\textit{nqreg}.
$$

Because the remaining gate list is empty
at the end of the loop 
(i.e., the loop condition ``$\texttt{remain.size() != 0}$'' becomes false at line~6 {\reviseg in Figure~\ref{fig;basicswap_code}}),
the \framework{} preprocessor will replace
the entire loop with an assumption  that
$[\![\texttt{output}]\!]_\textit{nqreg} \equiv [\![\texttt{input}]\!]_\textit{nqreg}$
and generate the following transformed code
that will be symbolically executed by Z3Py later.
\begin{lstlisting}[style=qubit]
def run(self, input):
    output = uninterpreted_circuit()
    assume(equivalent(output, input))
    return output
\end{lstlisting}

To validate the inferred loop invariant,
\framework{}  will   generate a separate proof goal that 
{\reviseg the invariant holds on} the symbolic execution results of the loop body.

\para{Expand branch statements} 
To support branches in the Z3Py tool,
\framework{} %
generates the sequential code 
for each branch. 
As for the  loop body in the
CXCancellation pass, three pieces of sequential code will be generated: 1) \texttt{gate} is a CNOT gate and a matching gate is found to perform
{\reviseg the} cancellation; 2) \texttt{gate} is a CNOT gate and no matching gates are found;
and 3) \texttt{gate} is not a CNOT gate. 
The branch conditions will be added as assumptions to the
generated sequential code.
{\reviseg For example,} the transformed code for the first branch is as follows:
\begin{lstlisting}[style=qubit]
    gate = remain[0]
    assume(gate.isCXGate())
    next = next_gate(remain, 0)
    g = remain[next]
    assume(g.isCXGate() and g.qubits == gate.qubits)
    remain.delete(next)
    remain.delete(0)
\end{lstlisting}

\para{Replace utility functions with specifications}
The \qiskit{} compiler implementations rely on many utility functions, 
which are shared by different passes and may contain code that is hard to 
automatically verify. 
\framework{} extracts a library of utility functions
and manually verifies all the functions with respect to their
specifications. %
A compiler pass can then be retrofitted to invoke utility functions in \framework{}'s verified library.
Such invocations will be replaced by the corresponding specifications
of the utility functions during the symbolic execution.
Take the \texttt{next\_gate} utility function 
{\reviseg that is used in four optimization passes of Qiskit} 
as an example.
This function scans through the circuit and returns the index of the next gate that shares a qubit with the current gate. 
Instead of repeatedly verifying this function whenever it is invoked,
\framework{} replaces its invocations with the verified specification.
For example,
its invocation at line~9 in {\reviseg Figure~\ref{fig;basicswap_code}}
will be replaced by the following specification
about the returned integer \texttt{x} of \texttt{next\_gate(remain, 0)}:
1) \texttt{x} is a valid index of the \texttt{remain} circuit, i.e., $0 \le\texttt{x} < \texttt{remain.size()}$; 2) \texttt{x} is after the gate $0$, i.e., \texttt{x} $> 0$; 3) there is no gate between gates $0$ and \texttt{x} that shares a qubit with gate $0$; and 4) gate \texttt{x} shares a qubit with gate $0$.

\para{Verify proof goals with rewrite rules} 
The \framework{} verifier {\reviseg first symbolically executes}
the transformed code  %
using the Z3Py tool. The Z3 SMT solver will then be queried to solve the proof goals
using the symbolic execution results.
As for the CXCancellation pass,
we need to prove that all three branches
will preserve the equivalence of circuits, i.e., 
    $[\![\texttt{output};\texttt{remain}]\!]_\textit{nqreg} \equiv [\![\texttt{input}]\!]_\textit{nqreg}$.
The only non-trivial case is 
the branch where a cancellation happens. 

Because the circuits are equivalent before executing the branch,
and the \texttt{input} and \texttt{output}   circuits
remain unchanged in this branch (since CNOT gates are cancelled out and will not be appended to \texttt{output}), 
we only need to prove that the \texttt{remain} list of gates
has the same denotational semantics 
after cancelling out two adjacent CNOT gates.
Such a proof goal can be stated 
using the following matrix representations:
$$
    \left(\texttt{matrix}(CNOT_{q_1, q_2})\otimes I_{\{1, \ldots, \textit{nqreg}\} \backslash \{q_1, q_2\}}\right)^2 = I_{2^{\textit{nqreg}}}
$$
For a quantum circuit of $n$ qubits, the above proof goal requires
the equivalence check of matrices with a size $2^n \times 2^n$, which is impractical to solve using any existing solvers.
The fact that $n$,  $q_1$, and $q_2$ are arbitrary makes such
a check even harder.

\framework{} introduces a set of {\reviseg rewrite} rules
to enable the equivalence check for quantum circuits
at the symbolic level
without the need
to reason about their denotational semantics.
These rules are also qubit-based, meaning that we only need to prove that the two related qubits ($q_1$ and $q_2$) are equivalent and do not need to worry about $n$ and the relative location of $q_1$ and $q_2$. 
For example, \framework{} provides the following two {\reviseg rewrite} rules %
to cancel out two adjacent CNOT gates,
with which the equivalence check for \texttt{remain}
can be solved automatically and efficiently 
by  Z3.
\begin{align*}
        \texttt{app}(CNOT\ q_1\ q_2; CNOT\ q_1\ q_2, q_1)  &\equiv q_1\\
       \texttt{app}(CNOT\ q_1\ q_2; CNOT\ q_1\ q_2, q_2)   &\equiv q_2.
\end{align*}

\section{The \framework{} Preprocessor}
\label{sec:\framework{}lang}
 The \framework{} preprocessor aims to soundly transform the compiler source code, with complex control flow and external function calls, into simple sequential code with verification conditions, thus symbolically executable and verifiable using Z3Py.
A verification condition is defined as a Hoare triple $\{P\}C\{Q\}$, with a pre-condition list $P$ and a post-condition list $Q$~\cite{Hoare1969}. 
{\reviseg The Hoare triple} means that if the program state before executing the code $C$ satisfies $P$, then after executing $C$, the resulting state must satisfy $Q$. \framework{} uses  \texttt{assume} and \texttt{assert} primitives
in Z3Py
to represent the pre- and 
post-conditions   respectively.
The \framework{} preprocessor parses and transforms the pass implementation
with complex control flows into sequential code and corresponding verification conditions.
 
\para{Branch statements}
The \framework{} preprocessor expands all branch statements.
The verification condition
of a branch is expanded
into two separate verification conditions---one for each branch.
The %
branch condition will be added to the list
of pre-conditions for the transformed sequential 
code representing the ``true'' branch, 
while the negation of the branch condition will be added to the list
of pre-conditions
 for the ``false'' branch. 
These pre-conditions are then used to generate further verification conditions. 
Note that \framework{} requires that all branch conditions must be representable as SMT formulas to enable
the push-button verification.
Although this expansion approach may lead to an exponential number of verification conditions, fortunately, the number of branches in real \qiskit{} compiler passes
is usually smaller than nine and remains tractable.

\para{Loop statements}
Unbounded loops are hard to automatically verify  
since their verification requires 
a sufficiently strong loop invariant,
which is generally undecidable to compute and usually provided by the user.
Fortunately, in the domain of quantum compilation,
this problem can be practically solved as loops in quantum compiler passes often follow one of a few fixed patterns. \framework{} provides \emph{three} loop templates as library functions, %
which
can be used to rewrite all the loops in the %
\qiskit{} compiler passes.
\framework{} then statically analyzes the loop implementation
and automatically determines 
how the placeholders in the loop template
can be mapped to the 
variables in the loop implementation.

Each of \framework{}'s loop template pre-defines one shape of the
loop invariants. 
For example,  \texttt{iterate\_all\_gates(circ, func)} is
a template for loops that iterate over all the gates in \texttt{circ} and apply the  same function \texttt{func} (i.e., the loop body) to each gate to generate the new circuit, denoted as
\texttt{new\_circ}.
Since this compiler pass  preserves the  semantics of the circuits,
the generated \texttt{new\_circ} should be equivalent to 
a prefix of
the original circuit.
Thus, the invariant for this loop template is that, %
at the $i$-th iteration of the loop,
\texttt{new\_circ} %
should be equivalent to the first $i$ gates in the original circuit. 
This loop template
is implemented as follows:
\begin{lstlisting}[style=qubit,numbers=left,xleftmargin=2em,framexleftmargin=1.5em]
def iterate_all_gates(circ, func):
    # subgoal
    assertion.push()
    i = Int("i")
    n = circ.size()
    cur_circ = QCircuit()
    assume(i >= 0)
    assume(i + 1 < n)
    assume(equiv_part(cur_circ, circ, i))
    new_circ = func(cur_circ, circ[i])
    assert(equiv_part(new_circ, circ, i+1))
    assertion.pop()
    
    ret_circ = DAG()
    assume(equiv_part(ret_circ, circ, circ.size()))
    return ret_circ
\end{lstlisting}
Lines~7-12 check
if the loop body (i.e., \texttt{func}) preserves the loop invariant.
In the verification process, when \framework{} invokes this loop template, it will generate and try to prove the subgoal that if the current %
circuit \texttt{cur\_circ}  is equivalent to the first $i$ gates of the original
circuit \texttt{circ} before the $i$-th iteration, the  newly generated circuit \texttt{new\_circ} after this iteration must be equivalent to the first $i+1$ gates of \texttt{circ}. 
Once this subgoal is proved,
the loop invariant is valid
and a new pre-condition stating that 
the result of the loop is a circuit that is equivalent to \texttt{circ}
will be added into the pre-condition list
(see line~15).
To synthesize the exact loop invariant
from the code,
\framework{} will  infer
the variable name in the loop body that should be mapped 
to the \texttt{circ} argument of 
\texttt{iterate\_all\_gates(circ, func)}.

The other two loop templates provided by \framework{} 
are  \texttt{while\_gate\_remaining}
and \texttt{collect\_runs}.
The unbounded loop 
in the \texttt{CXCancellation} pass (see {\reviseg Section~\ref{sec:framework}})
can be implemented using  the
\texttt{while\_gate\_remaining} template, which
maintains a remaining gate list to be scanned. 
The loop invariant for this template is that at each iteration of the loop,
the concatenation of the currently built part of output circuit and the remaining gates is equivalent to the input circuit. %
{\reviseg 
The 
\texttt{collect\_runs} template
is the batch version of the \texttt{iterate\_all\_gates} template
and 
is used in passes such as  \texttt{commutative\_cancellation} (see Section~\ref{sec:cases_comm}).
In this template, the} input circuit is partitioned into several batches
{\reviseg and each} loop round will transform one batch of the circuit into an equivalent circuit. The invariant for this template is that the 
{\reviseg currently} built output circuit in the $i$-th iteration of the loop is equivalent to the combination of first $i$ batches of the input circuit.

\para{Utility function calls} 
Many compiler passes are implemented using
{\reviseg some shared}
utility functions,
including circuit manipulating functions such as \texttt{next\_gate}%
, gate optimization functions such as \texttt{merge}, and coupling map related functions such as \texttt{shortest\_path}. 
To enable the push-button verification for \qiskit{} compiler passes
without the need to unfold and repeatedly verify these 
{\reviseg shared} utility functions
at all their invocations,
\framework{} pre-verifies all these functions with respect to their specifications %
and replaces their invocations with the specifications during the symbolic execution.
Take the \texttt{next\_gate}
utility function invoked
by four passes (CXCancellation, MergeAdjacentBarriers, RemoveFinalMeasure and RemoveDiagBeforeMeasure)
as an example.
\framework{} %
models a quantum gate as a record type with two fields---an operation name and a qubit list (analogous to the opcode and operands in classical computing)
and model a quantum circuit %
as a list of gates. \framework{} verifies
that the \texttt{next\_gate}
function meets the specification %
that scans through the circuit list and finds the index of the first gate that shares a qubit with the given gate. 
Although the library of utility functions  is verified manually, 
this verification effort is done once and for all
can be re-used in future \qiskit{} compiler passes.

Note that \framework{}'s verified utility library
implements a quantum circuit as a list of gates,
while the original \qiskit{} library implements a circuit
as a directed acyclic graph (DAG)
of gates.
\framework{}'s design  significantly simplifies the library's verification 
since lists are much easier to reason about in Coq than DAGs.
To integrate passes implemented using  
our verified library into the \qiskit{} platform,
\framework{} provides %
conversion functions to convert circuits implemented in different data structures,
as well as
a \qiskit{} wrapper  
that performs the following steps:
1) it first converts the input DAG circuit from the \qiskit{} compilation flow to the OpenQASM IR; 2) then it invokes {\revise the compiler pass written using \framework{}}
to 
compile the converted circuit represented as  a gate list;
and 3) it finally converts the compiled circuit back to the corresponding DAG representation. 
The evaluation section (see {\reviseg Section~\ref{sec:evaluation}}) later shows that
\framework{}'s verified library,  conversion functions,
and the \qiskit{} wrapper will only introduce
negligible performance overhead compared with the original implementation.

\begin{revise}
\para{Non-critical statements}
Many \qiskit{} compiler passes contain 
\emph{non-critical} statements and function invocations
that will not affect the generated circuits,
such as the functions that output %
the number of gates and the number of tensor factors in the circuit for the debugging purpose.
Take the NumTensorFactors pass which calculates the number of tensor factors in the circuit as an example.
These \emph{non-critical} statements will not 
affect our semantic-preserving proofs for the generated circuits and will be discarded during the pre-processing
before the symbolic execution.

\end{revise}

\section{Rewrite Rules for Quantum Circuits}\label{sec:equivalence}
Many of  the generated verification conditions  can be proved directly in Z3Py,
except for the
equivalence assertion {\reviseg stating} that two quantum circuits with variable lengths
are equivalent, i.e., 
given any input quantum state, the two circuits will produce the same output quantum state.
Checking the equivalence of quantum circuits using their matrix representations 
is intractable for Z3 
due to the exponential computational cost,
especially when the matrices have variable sizes.
This is a unique challenge for quantum computing and is crucial for 
automated verification of quantum compiler passes.
To address this challenge,
{\reviseg besides the symbolic execution of pass implementations,}
we introduce {\reviseg a new layer of symbolic representation and execution} 
for quantum circuits, as well as 
rules to reduce gates and perform circuit rewriting. 
Note that one should not confuse the symbolic {\reviseg execution} of 
quantum circuits described in this section with the symbolic execution of {\reviseg pass implementations provided by Z3Py}---the former takes quantum states as input/output while
the latter takes quantum circuits as input/output.

\para{Symbolic execution of quantum circuits}
A multi-qubit  quantum register $Q$ is symbolically represented as an array of symbolic qubits
$(q_1,\cdots,q_n)$.
\framework{} defines the
symbolic function
$\texttt{app}_{1q}(U, q)$
to denote the resulting qubit
of applying the 1-qubit gate
$U$ on $q$,
and defines $\texttt{app}_{2q}(U, q_1, q_2, k)$
to denote the $k$-th resulting qubit ($k\in\{1,2\}$) of applying the 2-qubit gate $U$
on $q_1$ and $q_2$. 
The result of applying the whole circuit $C$ to a symbolic quantum register is represented as \texttt{app}$(C, Q)$,
which can be executed by  applying each gate in sequence on  $Q$
{\reviseg and is defined as follows:}
{\small
\begin{align*}
    \texttt{app}(skip, Q) &\rightharpoonup Q \\
        \texttt{app}(C_1;C_2, Q) &\rightharpoonup  \texttt{app}(C_2, \texttt{app}(C_1, Q))\\
    \texttt{app}(U(i), (q_1,  \ldots, q_n)) &\rightharpoonup (q_1, \ldots, \texttt{app}_{1q}(U, q_i), \ldots,  q_n) \\
    \texttt{app}(U(i,j), (q_1,  \ldots, q_n)) &\rightharpoonup \\ (q_1, \ldots, \texttt{app}_{2q}(&U, q_i, 1), \ldots,\texttt{app}_{2q}(U, q_j, 2), \ldots,  q_n).
\end{align*}
}%
For example, applying the following GHZ  circuit 
\begin{center}\small
$GHZ :=\ H(0); CX(0, 1); CX(1, 2)$
\end{center}
on the register $(q_0, q_1, q_2)$
will result in $(q_0', q_1', q_2') $ such that
{\small
\begin{align*}
q_0' &= \texttt{app}_{2q}\left(CX, \texttt{app}_{1q}(H, q_0), q_1, 1\right) \\
q_1' &= \texttt{app}_{2q}\left(CX, \texttt{app}_{2q}(CX, \texttt{app}_{1q}(H, q_0), q_1, 2), q_2, 1\right) \\
q_2' &= \texttt{app}_{2q}\left(CX, \texttt{app}_{2q}(CX, \texttt{app}_{1q}(H, q_0), q_1, 2), q_2, 2\right).
\end{align*}
}%

  \begin{figure}[t]
     \centering \small

\begin{minipage}{0.30 \textwidth}
	\centering
    \Qcircuit @C=0.81em @R=0.1em {
	& \ctrl{2}    & \ctrl{2} & \qw 
	&&  &\qw&\qw&\qw    &&  &\qw&\qw&\qw&\qw\\
	&&&&\qceq& &&&  &\qceq\\
    & \targ       & \targ    & \qw
    &&  &\gate{H} &\gate{H} &\qw    &&  &\qw&\qw&\qw&\qw\\
	}
\end{minipage} \\
\vspace*{0.3cm}\noindent\rule{8.5cm}{0.6pt}\vspace*{0.3cm}

\begin{minipage}{0.10 \textwidth}
    \centering
    \Qcircuit @C=0.3em @R=0.2em{
&\gate{Z}    &\ctrl{2}  &\qw  &&&\ctrl{2} &\gate{Z}   &\qw\\
&&& &\qceq \\
&\qw         &\targ     &\qw  &&&\targ    &\qw &\qw\\
}
\end{minipage}\hspace*{0.8cm}
\begin{minipage}{0.10 \textwidth}
    \centering
    \Qcircuit @C=0.3em @R=0.2em{
&\gate{X}    &\targ  &\qw  &&&\targ &\gate{X}   &\qw\\
&&& &\qceq \\
&\qw         &\ctrl{-2}     &\qw  &&&\ctrl{-2}    &\qw &\qw\\
}
\end{minipage}\\\vspace*{0.3cm}
\begin{minipage}{0.10 \textwidth}
    \centering
    \Qcircuit @C=0.4em @R=0.7em{
&\targ      &\qw        &\qw    &&&
&\qw        &\targ      &\qw\\
&\ctrl{-1}  &\ctrl{1}   &\qw    &&\qceq&
&\ctrl{1}   &\ctrl{-1}  &\qw\\
&\qw        &\targ      &\qw    &&&
&\targ      &\qw        &\qw
}
\end{minipage}\hspace*{0.8cm}
\begin{minipage}{0.10 \textwidth}
    \centering
    \Qcircuit @C=0.4em @R=0.7em{
&\ctrl{1}   &\qw        &\qw    &&&
&\qw        &\ctrl{1}   &\qw\\
&\targ      &\targ      &\qw    &&\qceq&
&\targ      &\targ      &\qw\\
&\qw        &\ctrl{-1}  &\qw    &&&
&\ctrl{-1}  &\qw        &\qw
}
\end{minipage}\vspace*{0.3cm}\\ \noindent\rule{8.5cm}{0.6pt}\vspace*{0.3cm}\\
\begin{minipage}{0.30 \textwidth}
	\centering
    \Qcircuit @C=0.7em @R=0.7em {
	& \qswap        &\qw & \qswap           & \qw      &     & &\qw&\qw&     &\\
	&               &    &                  &          &\qceq& &   &   &\qceq& \\
    & \qswap\qwx[-2]&\qw & \qswap\qwx[-2]   & \qw      &     & &\qw&\qw&     &\\
	}
\end{minipage}
\begin{minipage}{0.10\textwidth}
	\centering
    \includegraphics[width=0.99\textwidth]{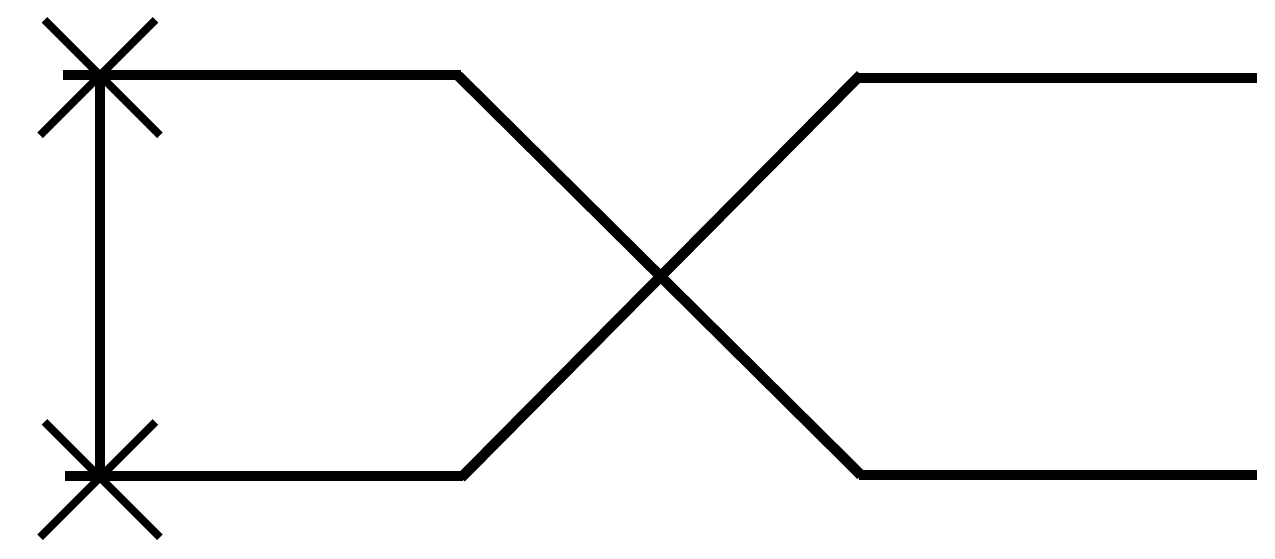}
\end{minipage}\\
    \vspace*{0.3cm}
     \caption{Examples of  rules for reducing and rewriting circuits. They are cancellation rules (above), commutativity rules (2nd, 3rd line), and swap rules (bottom). }
     \label{fig:primitive_moves}
         \vspace*{0.3cm}
 \end{figure}
 
\para{Rewrite rules}
 To efficiently check the equivalence
 of {\reviseg the symbolic execution}
 of quantum circuits,
 \framework{} introduces a set of 20
 rules
 for reducing and rewriting gates
 that are inserted
 by compiler passes, 
{\reviseg eight of which are shown}
in  
 {\reviseg Figure~\ref{fig:primitive_moves}}.
For example, 
a $SWAP$ gate will swap the two qubit arguments
and its application can be reduced by the following 
swap rules:
{\small
\begin{align*}
\texttt{app}_{2q}(SWAP, q_1, q_2, 1) &\equiv q_2 \\
\texttt{app}_{2q}(SWAP, q_1, q_2, 2) &\equiv q_1.
\end{align*}
}%

The cancellation rules say that applying two adjacent $CX$ gates on the same pair of qubits will not change the state:
{\small
\begin{align*}
    &\texttt{app}_{2q}(CX, \texttt{app}_{2q}(CX, q_1, q_2, 1), \texttt{app}_{2q}(CX, q_1, q_2, 2), 1) \equiv q_1 \\
    &\texttt{app}_{2q}(CX, \texttt{app}_{2q}(CX, q_1, q_2, 1), \texttt{app}_{2q}(CX, q_1, q_2, 2), 2) \equiv q_2.
\end{align*}
}%

These {\reviseg rewrite} rules are
defined as Z3 \texttt{assertions} 
about the symbolic functions 
$\texttt{app}_{1q}$ and $\texttt{app}_{2q}$,
and are added into the  precondition list of the proof obligation.

\para{Soundness proofs}
The symbolic {\reviseg execution and
rewrite rules} for quantum circuits
are treated as axioms in \framework{}
and are proven using the denotational semantics
defined using
the QWire matrix library~\cite{qwire}
within the Coq proof assistant~\cite{Coq12}. 
{\reviseg The soundness of symbolic execution of quantum circuits}
can be trivially verified by showing that
the output state  always has the same
 denotational semantics
with the input state.
Since all {\reviseg  rewrite} rules
only deal with a small number of gates,
their soundness   can also be proven easily using
the matrix representation.
We also prove in Coq that the equivalence guarantee of the rewrite rules on a subset of qubits
can be extended to the global circuit with an arbitrary number of qubits.
Again, %
although these soundness proofs are developed manually in Coq, 
they only need to be done once.

\section{The \framework{} Verifier}
\label{sec:verifier}
The \framework{} verifier generates proof obligations
for  compiler passes
as SMT problems and invokes
Z3 to solve them
with the {\reviseg rewrite} rules.

\para{Proof obligations for a compiler pass}
\framework{} does not require the user to explicitly 
provide the specifications to the compiler pass,
which is usually required by other automated verification frameworks
such as Yggdrasil~\cite{yggdrasil} {\reviseg and Serval~\cite{serval}}.
Instead, \framework{} pre-defines  %
the proof obligations for all seven types of quantum compiler passes.
Aside from routing passes,
the other six types of compiler passes %
share the same specification that the input and output circuits are equivalent,
defined using
the following virtual class:
\begin{center}
\begin{tabular}{c}
\begin{lstlisting}[style=qubit]
class GeneralPass():
    @classmethod
    def test(cls):
        optimizer = cls()
        init_circ = QCircuit()
        out_circ = optimizer.run(init_circ)
        assert(equivalent(out_circ, init_circ))
        print(cls.__name__ + " verified")
\end{lstlisting}
\end{tabular}
\end{center}

\noindent{}In the above \texttt{test()} method, \framework{} first generates a symbolic circuit \texttt{init\_circ} to represent the input
quantum circuit, then symbolically executes the pass implementation
through \texttt{optimizer.run} to get the symbolic representation of the output circuit,
and finally attempts to verify that these two circuits are equivalent.

To enable the automated generation of proof
obligations,
\framework{} requires the user to retrofit the compiler passes
(in the above six types) with
the \texttt{GeneralPass} virtual class as the parent class.
{\reviseg For example, the class definition for the CXCancellation
pass is shown as follows:}
\begin{center}
\begin{tabular}{c}
\begin{lstlisting}[style=qubit]
class CXCancellationPass(GeneralPass):
    #implementation omitted
    #...
\end{lstlisting}
\end{tabular}
\end{center}

Different from the other six types,
the routing passes may insert swap gates
to make the circuit satisfy the 
qubit connectivity constraint (see {\reviseg Section~\ref{sec:qcompilation}}), such that
the output circuit may not be strictly equivalent to the input circuit. 
However, the output circuit
is equivalent to the input circuit up to a permutation that represents 
all inserted swaps.
\framework{} provides a \texttt{RoutingPass} virtual class 
for users to implement routing passes, whose
proof obligations will then be
 automatically generated by \framework{}.

\para{Verification for a compiler pass}
We will use the CXCancellation pass (see {\reviseg Section~\ref{sec:framework}}) as {\reviseg a running} example
to show how the \framework{} verifier works.
This CXCancellation pass
is an optimization pass
and contains an unbounded loop
maintaining two circuit variables \texttt{output} and \texttt{remain}. 
The variable \texttt{output} contains gates that have been scanned, while 
\texttt{remain} contains gates that have not been scanned yet. 
Each loop iteration will attempt to find two CNOT gates with
the same pair of input qubits in \texttt{remain} and cancel them out.

The CXCancellation pass inherits the \texttt{GeneralPass}  virtual class
and the  proof obligation generated by \framework{} %
is to show
that the input and output circuits are equivalent,
which can be derived using the 
\texttt{while\_gate\_remaining} loop template.
\framework{} will then attempt to solve the following subgoal
generated for the loop body:
\begin{align*}
{\bf P_1}:\quad  &
\texttt{app}\left(\texttt{output}_{old};\texttt{remain}_{old}, Q\right) \equiv \texttt{app}\left(\texttt{input}, Q\right) \\
{\bf G_1}:\quad  &
\texttt{app}\left(\texttt{output}_{new};\texttt{remain}_{new}, Q\right) \equiv \texttt{app}\left(\texttt{input}, Q\right), 
\end{align*}
where the new symbolic values for variables \texttt{output} and \texttt{remain}  
are produced by the loop body from the old ones.
{\reviseg
To generate the relation between the old and new variables,
}
the \framework{} verifier uses Z3Py to symbolically execute the loop body, getting the following preconditions:
\begin{align*}
 {\bf P_2}:\quad &   \texttt{output}_{new} = \texttt{output}_{old} \\
 {\bf P_3}:\quad    & \texttt{remain}_{old}\texttt{[0]} = CX \\
 {\bf P_4}:\quad &    \texttt{remain}_{old}\texttt{[x]} = CX \\
  {\bf P_5}:\quad &   \texttt{remain}_{new} = \texttt{remain}_{old}\texttt{[1:x]};\texttt{remain}_{old}\texttt{[x+1:]},
\end{align*}
where \texttt{x} is the return value of \texttt{next\_gate(remain, 0)}. For convenience, we use $C_1$ and $C_2$ to represent $ \texttt{remain}_{old}\texttt{[1:x]}$ and $\texttt{remain}_{old}\texttt{[x+1:]}$ respectively, and thus we have \begin{align*}
& \texttt{remain}_{old} = CX; C_1; CX; C_2 \\
& \texttt{remain}_{new} = C_1; C_2.
\end{align*}
Using ${\bf P_1} \sim {\bf P_5}$, we can rewrite the
proof goal ${\bf G_1}$ as:
$${\bf G_2}:\quad  
\texttt{app}\left(CX; C_1; CX; C_2, Q'\right) \equiv \texttt{app}\left(C_1;C_2, Q'\right),$$ 
{\reviseg where $Q'$ is the quantum state after applying 
$\texttt{output}_{old}$ or $\texttt{output}_{new}$ to $Q$.}
 \framework{}  then performs symbolic execution on both sides of ${\bf G_2}$, {\reviseg shown as follows:}
{\small
\begin{align*}
\textbf{Left: }\quad &  \texttt{app}(CX;C_1;CX;C_2, Q')\\
&\rightharpoonup \texttt{app}\left(C_1;CX;C_2, \texttt{app}(CX,Q')\right)\\
&\cdots\\
&\rightharpoonup \texttt{app}\left(C_2, \texttt{app}(CX, \texttt{app}(C_1, \texttt{app}(CX,Q')))\right) \\[5pt]
\textbf{Right: }\quad &  \texttt{app}(C_1;C_2, Q') \rightharpoonup 
\texttt{app}(C_2, \texttt{app}(C_1, Q')).
\end{align*}
}%
\noindent{}Thus, after removing the same  application of $C_2$ on both sides
of ${\bf G_2}$, the proof goal becomes:
$${\bf G_3}:\quad  
\texttt{app}\left(CX, \texttt{app}(C_1, \texttt{app}(CX,Q'))\right)\equiv \texttt{app}(C_1, Q').$$ 

Note that the specification of \texttt{next\_gate(0)} also adds the following precondition
that the gate \texttt{x} can be reordered to the second gate of $\texttt{remain}_{old}$:
\begin{align*}
  {\bf P_6}: \quad \forall Q_1, \texttt{app}\left(CX, \texttt{app}(C_1, Q_1)\right) \equiv \texttt{app}\left(C_1, \texttt{app}(CX, Q_1)\right).
\end{align*}

Besides, \framework{}'s {\reviseg rewrite rule set contains}
a cancellation rule to
cancel out two adjacent $CX$ gates,
which is introduced as the following precondition to the proof goal:
\begin{center}
${\bf P_7}: \quad
\forall Q_2,\ \texttt{app}\left(CX, \texttt{app}(CX, Q_2)\right) \equiv Q_2
$ 
\end{center}

The \framework{} verifier
{\reviseg finally} encodes the preconditions
and  goals  into the following formula
and invokes Z3 to solve:
\begin{center}
${\bf P_6} \wedge {\bf P_7} \wedge \neg {\bf G_3},
$ 
\end{center}
\noindent{}which is expanded into qubit-local formulas containing $\texttt{app}_{2q}$ of symbolic representation for quantum circuits (see Section~\ref{sec:equivalence}).
If the above formula
is satisfiable,
the compiler pass is incorrect
and a counter-example is generated by the verifier. Otherwise when Z3 proves the above formula
is unsatisfiable, 
we successfully verify that the compiler pass  
correctly preserves the semantics using \framework{}.

\begin{reviseg}
The above formula is unsatisfiable, i.e., ${\bf P_6}$ and ${\bf P_7}$  imply ${\bf G_3}$, and can be proven using Z3.
We show why ${\bf P_6}$ and ${\bf P_7}$  imply ${\bf G_3}$ as follows.
By instantiating $Q_1$ with $\texttt{app}(CX,Q')$
and rewriting the left side of ${\bf G_3}$ using ${\bf P_6}$, the proof goal becomes:
$${\bf G_4}:\quad  
\texttt{app}\left(C_1, \texttt{app}(CX, \texttt{app}(CX,Q'))\right)\equiv \texttt{app}(C_1, Q').$$ 
After removing the same  application of $C_1$ on both sides
of ${\bf G_4}$, the proof goal becomes:
$${\bf G_5}:\quad  
\texttt{app}\left(CX, \texttt{app}(CX,Q')\right)\equiv Q',$$ 
which can be directly proven using the cancellation
rule  ${\bf P_7}$.
\end{reviseg}

\begin{revise}
\para{Requirements of the \framework{} verifier} For a pass to be verified using \framework{}, it needs to be written in a way that 1) each loop in the pass follows one of the three loop patterns introduced in Section~\ref{sec:\framework{}lang};  2) it uses \framework{}'s verified library to represent quantum programs;
and 3) the transformation of quantum gates must be expressible using the given rewrite rules.
\end{revise}
{\reviseg Note that new loop templates, verified library functions,
and rewrite rules can be easily added.}

\section{Case Studies}
\label{sec:cases}

In this section,
we will present three case studies to show how we use \framework{} to discover quantum-specific bugs during the push-button verification of the \qiskit{} compiler.

 \subsection{The \texttt{optimize\_1q\_gate} Pass}
 \label{sec:op_1q_pass}
We first focus on the verification of the \texttt{optimize\_1q\_gate} pass and show that, using \framework{}, we can reveal bugs that only arise in quantum software.

The \texttt{optimize\_1q\_gate} pass invokes the utility function \texttt{merge\_1q\_gate}
to collapse a chain of 1-qubit gates into a single, more efficient gate~\cite{McKay2018}, to mitigate noise accumulation. It operates on $u_1$, $u_2$,
and $u_3$ gates, which are native gates in the IBM quantum devices. %
These gates can be naturally described as linear operations on the Bloch sphere; for example, $u_1$ gates are rotations with respect to the Z axis. For clarity, we list their matrix representations in Table~\ref{tab:gates}. 

\begin{table}[t] 
    \centering \small
    \caption{Matrix representation of physical gates $u_1$, $u_2$ and $u_3$, where $u_1$ is a Z rotation on the Bloch sphere.}
    \begin{tabular}{l  c  r}
    $u_1(\lambda) = \begin{pmatrix}1&0\\0&e^{i\lambda}\end{pmatrix}$, &  \multicolumn{2}{c}{$u_2(\phi, \lambda) = \frac{\sqrt{2}}{2} \begin{pmatrix}1&-e^{i \lambda}\\e^{i \phi}&e^{i(\lambda+\phi)}\end{pmatrix}$} \\[0.55cm] \hline \\
    \multicolumn{3}{c}{$u_3(\theta, \phi, \lambda) = \frac{\sqrt{2}}{2} \begin{pmatrix}\text{cos}(\theta)&-e^{i \lambda}\text{sin}(\theta)\\e^{i \phi}\text{sin}(\theta)&e^{i(\lambda+\phi)}\text{cos}(\theta)\end{pmatrix}$}\\[0.55cm] 
    \end{tabular}
    \label{tab:gates}

\end{table}

The \texttt{optimize\_1q\_gate} pass has two function calls. First, it calls the \texttt{collect\_runs} method to collect groups of consecutive $u_1$, $u_2$, and $u_3$ gates. Then it calls \texttt{merge\_1q\_gate} to merge the gates in each group. The \texttt{merge\_1q\_gate} method (see \cref{fig:correct_mlg}) 
first transforms the 1-qubit gates from the Bloch sphere representation to the unit quaternion representation~\cite{Gille2009}
and then applies the \texttt{merge()} function to that %
representation to merge the rotations. 

In \qiskit{}, all gates can be modified with a \texttt{c\_if} or \texttt{q\_if} method to condition its execution on the state of other classical or quantum bits. 
When proving that \texttt{merge()} does not change semantics of the quantum program, we found that
in some cases the compiler pass attempts to optimize these gates without noticing that the gate is controlled by other (qu)bits, leading to an incorrect execution as in Figure \ref{fig:incorrect_mlg}. For this reason, in our retrofitted implementation of this pass, we inserted checks that \texttt{gate1.q\_if == False}  and \texttt{gate1.c\_if == False} before merging the 1-qubit gates. 
 
  Bugs similar to the one described above, which relates to how quantum circuit instructions can be conditioned, have been observed in \qiskit{} in the past~\cite{qiskitbug1,terra_issue}. In the absence of the rigorous verification provided by tools like Giallar, such bugs are  hard to discover. In practice, this is usually done via extensive randomized testing of input and output circuits, which does not provide any guarantee of finding faulty code.
 The results of \texttt{merge\_1q\_gate} here demonstrate that \framework{}  is effective for detecting quantum-related bugs. 
\begin{figure}[t]
    \centering \small
    \begin{subfigure}[b]{.8\columnwidth}
    \Qcircuit @C=.6em @R=.3em @!R {
    & \gate{u_1(\lambda_1)} & \gate{u_3(\theta_2, \phi_2, \lambda_2)} & \qw && &\xrightarrow{\texttt{\lstinline{m1g}}}&&&
    \gate{u_3(\theta_2, \lambda_1+\phi_2, \lambda_2)} & \qw
    }
    \caption {\label{fig:correct_mlg} A correct merge of gates $u_1$ and $u_3$. The cuicuit is equivalent before and after merging.}
    \vspace{.4cm}
    \end{subfigure} 
    \begin{subfigure}[b]{.8\columnwidth}
    \Qcircuit @C=.6em @R=.3em @!R {
    &\qw &\ctrl{1}&\qw
    &&&&&
    &\qw&\qw\\
     & \gate{u_1(\lambda_1)} & \gate{u_3(\theta_2, \phi_2, \lambda_2)} & \qw && &\xrightarrow{\texttt{\lstinline{m1g}}}&&&
    \gate{u_3(\theta_2, \lambda_1+\phi_2, \lambda_2)} & \qw    } 
    \caption{\label{fig:incorrect_mlg} An incorrect merge. The new curcuit is not equivalent because $u_3$ was a controlled gate before merging.}
    \end{subfigure}
    \vspace*{0.15cm}
      \caption{Correct execution (top) and incorrect execution  (bottom) of \texttt{\lstinline{merge\_1q\_gate}}.
      }
    \label{fig:opt1qpass}
\end{figure}

\subsection{The \texttt{commutation} Passes}
\label{sec:cases_comm}
The second bug found are in 
the \texttt{commutation\_\-analysis} and \texttt{commutative\_\-cancellation} 
passes which optimize \qiskit{} DAGCircuits using the quantum commutation rules and the cancellation rules pictured in 
{\reviseg Figure~\ref{fig:primitive_moves}}. First, \texttt{commutation\_\-analysis} transforms the quantum circuit to a representation called commutation groups~\cite{Shi2019}, where nearby gates that commute with each other are grouped together. Next, \texttt{commutative\_\-cancellation} performs cancellation inside the newly formed groups. We give a working example in {\reviseg Figure~\ref{fig:comm_pass}}.
The circuit in {\reviseg Figure~\ref{fig:comm_pass}a} is first partitioned into three parts such that all gates in one part commute with each other. For example, in the middle part in {\reviseg Figure~\ref{fig:comm_pass}b}, the $Z$ gate on qubit $0$ commutes with the following $CNOT$ on qubits $0, 1$, and the $X$ gate on qubit $1$ also commutes with $CNOT\ 0, 1$. These commutativity relations can be verified in Coq using matrix semantics and included %
in the {\reviseg rewrite} rules as shown in {\reviseg Figure~\ref{fig:primitive_moves}}.
\begin{figure}[t]
    \centering \small
    \vspace*{0.4cm}
    \hspace*{-0.2cm}
    \begin{minipage}{0.30 \textwidth}
    \centering
\Qcircuit @C=.5em @R=.5em @!R{
    &\targ  &\gate{Z} &\ctrl{1} &\gate{Z} &\ctrl{1} &\targ &\qw \\
    &\ctrl{-1} &\gate{X} &\targ &\qw &\targ &\ctrl{-1} &\qw
    }\vspace*{0.4cm} \hspace*{-2cm}(a) 
    \end{minipage}\hspace*{-2.0cm}
    \begin{minipage}{0.30 \textwidth}
    \centering
        \Qcircuit @C=.5em @R=.5em @!R{
    0\quad &\targ  &\qw&\gate{Z} &\ctrl{1} &\gate{Z} &\ctrl{1} &\qw&\targ &\qw \\
    1\quad &\ctrl{-1}&\qw &\gate{X} &\targ &\qw &\targ &\qw&\ctrl{-1} &\qw
    \gategroup{1}{2}{2}{2}{.7em}{--}
    \gategroup{1}{4}{2}{7}{.7em}{--}
    \gategroup{1}{9}{2}{9}{.7em}{--}
    }\vspace*{0.2cm}\hspace*{-2cm} (b) 
    \end{minipage}\hspace*{-1.7cm}
    \begin{minipage}{0.30 \textwidth}
    \centering
    \Qcircuit @C=.5em @R=.5em @!R{
    &\targ &\qw &\targ &\qw \\
    &\ctrl{-1} &\gate{X} &\ctrl{-1} &\qw
    }\vspace*{0.45cm}\hspace*{-3.5cm} (c) 
    \end{minipage}\\\vspace*{0.1cm}
    \caption{A working example of  \texttt{commutation\_\-analysis} and \texttt{commutative\_\-cancellation}. (a) The un-optimized circuit, (b) \texttt{commutation\_\-analysis} forming the commutation groups, and (c) \texttt{commutative\_cancellation} cancels out self-inverse gates within each group.}
    \label{fig:comm_pass}
   
\end{figure}

A bug was discovered when we attempted to verify these two passes. In the original implementation, gates in one group are not guaranteed to be pairwise commutative. We found this violation comes from the fact that the commutation relation is in general not transitive. For example, denoting the commutation relation as $\sim$, if there are three quantum gates, $A, B, C$ where $A\sim B$ and $B\sim C$, then $A\sim C$ is not guaranteed to be true. For this reason, gates that do not commute will be grouped together in the flawed version of this pass. In our fixed version, we make sure that the circuits compiled by these passes %
have a limited gate set where $\sim$ is indeed transitive. For example, in the gate set \{CX, X, Z, H, T, $u_1, u_2,u_3$\}, $\sim$ is transitive. %

\subsection{Termination in \texttt{routing} Passes}
\label{sec:cases_swap}

For passes with loops, besides proving that the output circuit is equivalent to the input circuit, we also need to prove that the pass will terminate. Among our three loop templates, \texttt{iterate\_all\_gates} and \texttt{collect\_runs} are  range-based  for-loops
and always terminate. The \texttt{while\_gates\_remaining} template is a while-loop %
and may not terminate. For the \texttt{while\_gates\_remaining} template, 
\framework{} will generate a subgoal to prove that after each iteration, the number of remaining gates strictly decreases, i.e., at least one gate in the remaining gate list is removed when executing the loop body. Most of the passes always call the \texttt{QCircuit.delete()} method at least once in the loop body, and thus they always terminate. However, in one pass, namely the \texttt{look\_ahead} routing pass, \framework{} fails to prove its termination.

\begin{figure}[t]
    \centering
    \begin{subfigure}[b]{0.4 \textwidth}
    \includegraphics[width=1.0\textwidth]{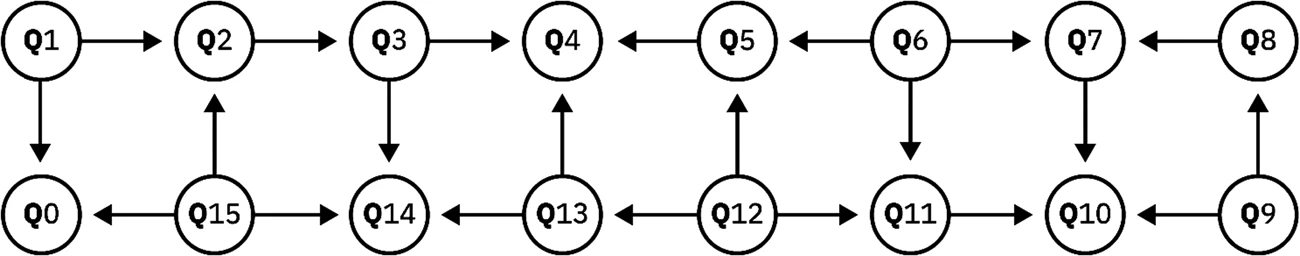}
    \caption{\label{fig:counter:a}}
    \end{subfigure}
        \begin{subfigure}[b]{0.4 \textwidth}\hspace{0.31\textwidth}
    \Qcircuit @C=0.8em @R=0.3em{
  &\lstick{\scriptstyle{Q_0}}  &\ctrl{1} &\qw        &\ctrl{3}   &\qw\\
  &\lstick{\scriptstyle{Q_8}}  &\targ    &\ctrl{1}   &\qw        &\qw\\
  &\lstick{\scriptstyle{Q_7}}  &\ctrl{1} &\targ      &\qw        &\qw\\
  &\lstick{\scriptstyle{Q_{15}}} &\targ    &\qw        &\targ      &\qw
    }\caption{\label{fig:counter:b}}
    \end{subfigure}
    \caption{(a) The coupling map of the IBM 16 qubit device. Arrows indicate available CNOT directions (which does not affect the swap insertion step). (b) A counter-example generated by \framework{} that shows \qiskit{}'s \texttt{lookahead\_swap} pass does not always terminate on the IBM 16 qubit device.
    }
    \label{fig:counter}
\end{figure}

After closer inspection of the pass, we found a counter-example circuit on the coupling map of the IBM 16 qubit device, for which the \texttt{lookahead\_swap} pass does not terminate (see {\reviseg Figure~\ref{fig:counter}}).
\begin{revise}
As shown in the coupling map of the IBM 16 qubit device (see Figure~\ref{fig:counter:a}),
a 2-qubit gate may not be available for two physical qubits.
The \texttt{lookahead\_swap} pass is designed to insert swap gates 
such that, for every 2-qubit gate, its two logical qubits are mapped
to physical qubits  allowing that gate.
The previous \texttt{lookahead\_swap} pass implementation
repeatly finds one \texttt{swap} gate to minimize the total distance of 
all 2-qubit gates at each iteration until all gates are mapped.
In the counter-example, if the four logical qubits is mapped to physical qubits $Q_0$, $Q_8$, $Q_7$ and $Q_{15}$ (see Figure~\ref{fig:counter:b}),
then the total distance will not change by inserting any single swap gate.
In this special case, the pass implementation will 
insert a swap gate between $Q_1$ and $Q_2$.
However, the total distance of the resulting map still will not change
by inserting any single swap gate. Thus, another swap gate will be inserted between $Q_1$ and $Q_2$. 
By applying the swap rules as in {\reviseg Figure~\ref{fig:primitive_moves}},
these two swap gates will cancel out and  the \texttt{lookahead\_swap} pass
will restart from the initial state and will not terminate.
\end{revise}

We fixed the bug by inserting the swap gate at a \emph{random} location
when the total distance will not change by inserting any single swap gate.
The randomization introduced by this fix will avoid keeping inserting swap gates
at the same location that can cancel out.
Because the choice of \texttt{swap} gates does not affect the correctness of the pass,
such randomness does not prevent the correctness verification. 
We proved the correctness of this randomized version by modeling random numbers as arbitrary symbolic numbers.

\section{Evaluation}
\label{sec:evaluation}
To demonstrate its effectiveness at verifying quantum compiler passes, we implemented and evaluated \framework{} on \totalPassVerified{} compiler passes in 
13 versions (from v0.19 to v0.32) of the \qiskit{} compiler
that are listed on the \qiskit{} website~\cite{qiskit_docs}.
The implementation consists of 3.6k lines of code for 
the \framework{} framework, including
quantum circuit-related data structures,
loop templates,
virtual classes,
the \framework{} preprocessor,
and the \qiskit{} wrapper,
and 168 lines of code for  utility functions.

Using \framework{}, we have successfully 
verified \numPassVerified{} out of the \totalPassVerified{} compiler passes.
Among all 12 passes that \framework{} fails to verify, 
eight passes are scheduling passes that deal with pulse instructions which are at a lower abstraction level than quantum gates.
Same as the previous verification frameworks for quantum computing~\cite{voqc,Amy2019},
\framework{} only supports the abstraction at the quantum gate level
and cannot reason about pulse-level behaviors.
The other four passes that we do not verify  are {\it StochasticSwap},  {\it CrosstalkAdaptiveSchedule}, {\it BIPMapping}, and {\it UnitarySynthesis} passes.
{\it StochasticSwap} uses a randomized routing algorithm that \framework{} does not yet support. 
{\it CrosstalkAdaptiveSchedule} and 
{\it BIPMapping}
are passes 
that %
invoke Z3 and CPLEX solvers respectively to find the output circuit. \framework{} does not have formal semantics for solvers
{\reviseg and therefore cannot}  model these two passes.
{\it UnitarySynthesis} is a synthesis pass that produces
an approximated circuit when the exact circuit includes gates that cannot be performed on the given quantum hardware.
{\reviseg This cannot be verified without a proper way to specify
and reason about the error bound of the approximated compilation.}
Note that all the above failed passes
are {\reviseg also} costly or infeasible to manually verify   
using previous verification frameworks for quantum compilers~\cite{voqc,Amy2019}.

\para{Experiment setup}
Our evaluation only focuses %
on %
the \numPassVerified{} verified passes. We have evaluated  the performance of verifying these \numPassVerified{} passes, as well as running verified passes to compile real quantum circuits compared with the unverified Qiskit passes.
The system ran with Python 3.8.7 and Z3 4.8.12. All verification and compilation tasks ran on a Dell Precision 5829 workstation with a 4.3GHz 28-core Intel Xeon W-2175, 62GB RAM and
a 512GB Intel SSD Pro 600p.

\para{Verification performance}
Table~\ref{tab:verification} gives the result of the verification 
{\reviseg for} all \numPassVerified{} Qiskit passes.
The verification of all passes completed in less than 30 seconds. 

Table~\ref{tab:verification} also lists the number of subgoals to be proved after preprocessing each pass. We can see that even if theoretically there may be an exponential number of subgoals when there are many branch statements, {\reviseg there are at most eight subgoals for all}  \qiskit{} passes.

\begin{table}[t]
    \centering\small
        \caption{Verification result of the \numPassVerified{} verified passes in \qiskit{}. %
    }
\begin{tabular}{c|cccc}
\hline
    
    \textbf{Pass} & \textbf{Pass} & \textbf{\#sub-} & \textbf{Verif.}  \\
 \textbf{name} & \textbf{LOC} & \textbf{goals} & \textbf{time(s)}\\
    \hline
    ApplyLayout & 11 & 2 & 0.7\\
    SetLayout & 8  & 1 & 0.7\\
    TrivialLayout  & 10  & 1 & 0.7\\
    Layout2qDistance  & 19  & 1 & 0.7\\
    DenseLayout  & 77  & 1 & 0.7\\
    NoiseAdaptiveLayout & 192 & 1 & 0.7\\
    SabreLayout & 62 & 1 & 0.7\\
    CSPLayout & 52 & 1 & 0.7\\
    EnlargeWithAncilla & 8 & 1 & 0.8\\
    FullAncillaAllocation & 8 & 1 & 0.7\\
    BasicSwap & 36 & 4 & 2.4 \\
    LookaheadSwap & 100 & 3 & 3.5\\
    SabreSwap & 96 & 3 & 3.8 \\
    Unroller & 23 & 3 & 1.5\\
    Unroll3qOrMore & 23 & 3 & 1.4\\
    Decompose & 23 & 3 & 2.0 \\ 
    UnrollCustomDefinitions & 22 & 3 & 1.5\\
    BasisTranslator & 119 & 3 & 1.5 \\
    Optimize1qGates & 32 & 3 & 25.1\\
    Optimize1qGatesDecomp & 32 & 3 & 25.2\\
    Collect2qBlocks & 9 & 1 & 0.7\\
    ConsolidateBlocks & 19 & 3 & 1.4\\
    CXCancellation & 24  & 4 & 4.2\\
    CommutationAnalysis & 6 & 1 & 0.7\\
    CommutativeCancellation & 17 & 3 & 1.3\\
    RemoveDiagBeforeMeasure & 24 & 3 & 18.1\\
    RemoveReseatInZeroState & 16 & 3 & 1.3\\
    Width & 8  & 1 & 0.6\\
    Depth & 8  & 1 & 0.6\\
    Size & 9  & 1 & 0.6\\
    CountOps & 8  & 1 & 0.7\\
    CoutOpsLongestPath & 8  & 1 & 0.7\\
    NumTensorFactors & 8  & 1 & 0.7\\
    DAGLongestPath & 8 & 1 & 0.9\\
    CheckMap & 19  & 1 & 0.7\\
    CheckCXDirection & 19 & 1 & 0.7\\
    CheckGateDirection & 19 & 1 & 0.7\\
    CXDirection & 29 & 4 & 2.3\\
    GateDirection & 55 & 8 & 5.6\\
    MergeAdjacentBarriers & 24 & 4 & 4.1\\
    BarrierBeforeFinalMeasure & 22 & 4 & 19.7 \\
    RemoveFinalMeasure & 20 & 3 & 2.9 \\
    DAGFixedPoint & 17 & 1 & 0.6 \\
    FixedPoint & 17 & 1 & 0.6\\
    \hline
    \textbf{Sum} & 1,366 & 95 & 145.4 \\
    \hline
\end{tabular}
    \label{tab:verification}
\end{table}

\para{Compiler performance}
We  have also evaluated our verified compiler and the original unverified \qiskit{} compiler on a series of quantum circuits from QASMBench~\cite{qasmbench}. The benchmark includes 48 quantum circuits with various near-term quantum applications, including quantum state preparation (cat\_state, bell, ghz\_state), quantum arithmetic (adder), quantum chemistry simulation (ising), quantum machine learning (dnn), and other famous quantum algorithms (deutsch, qft, grover, qaoa).
These quantum circuits contain up to 27 qubits and up to 5,000 quantum gates.

\begin{figure*}[ht]
    \centering
    \vspace{-5pt}
    \includegraphics[width=\textwidth]{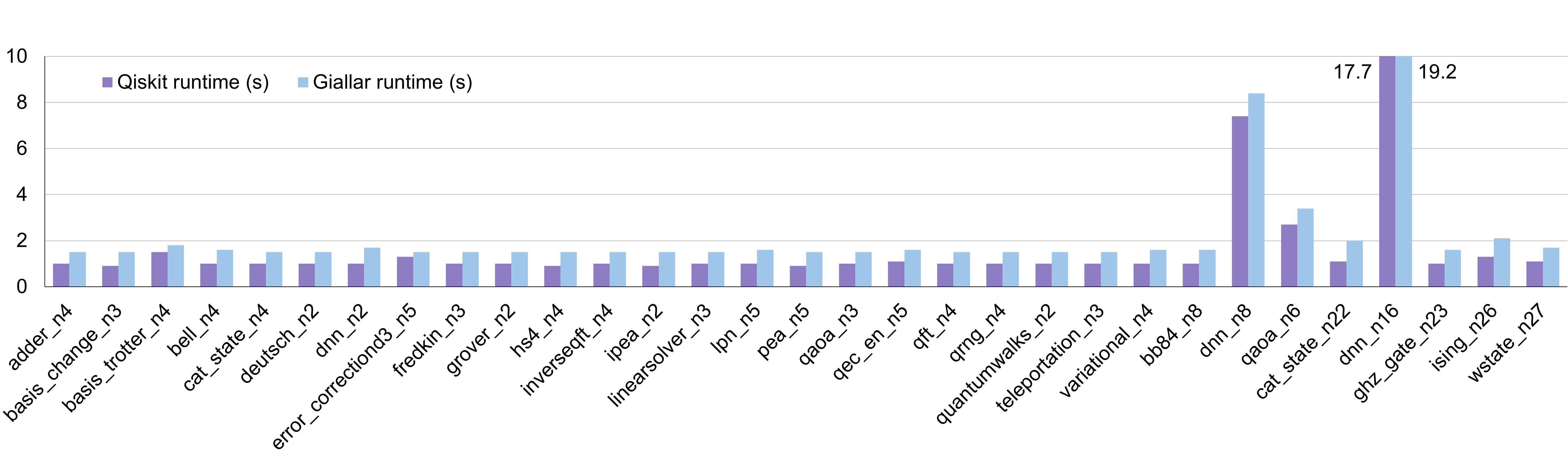}
    \vspace{-5pt}
    \caption{Comparison of the compilation performance of \qiskit{} and \framework{} on QASMBench.}
    \label{fig:benchmark}
\end{figure*}

We ran all 48 circuits in the benchmark using the (most computationally expensive) lookahead swap pass for the \qiskit{} implementation and the \framework{} implementation. Figure~\ref{fig:benchmark} summarizes the running time of all 31 benchmark circuits that \qiskit{} succeeded. We can see that \framework{} successfully compiled all these 31 circuits as well. %
For smaller circuits, the performance overhead of \framework{} was %
at most 0.5 seconds, while
for larger circuits, the performance overhead was at most 10\%. 
The overhead mainly came from loading the verified library in Python, the data structure conversion between \qiskit{} DAG and \framework{} list representations,
and the \qiskit{} wrapper.
The results show that the formal correctness guarantee of \framework{} 
{\reviseg  introduces only a modest compilation performance overhead.}

\begin{revise}
\para{Reusability}
Most of \framework{}'s rewrite rules and utility functions are shared across passes.
Among all three classes of rewrite rules shown in Figure~\ref{fig:primitive_moves}, the cancellation rules are used by optimization passes including CommutativeCancellation, CXCancellation, Optimize1qGates, and ConsolidateBlocks. The commutativity rules are used in the CommutativeAnalysis and CommutitiveCancellation passes. The swap rules are used in all routing passes. 
The utility functions in \framework{} include: 1) the circuit manipulating operations (e.g., \texttt{next\_gate}) which are used among all passes; 2) the coupling map related operations (e.g., \texttt{shortest\_path}) that are used in all routing passes; and 3) the circuit merging and gate expanding operations \texttt{merge} which are used in the ConsolidateBlocks and Optimize1qGate passes.
\end{revise}

\begin{reviseg}
\para{Adding new passes} Our experience shows that many new passes introduced can be verified automatically. \framework{} was first developed for Qiskit 0.19 (see our technical report~\cite{shi2019certiq}). When we applied Giallar to verify the 16 new passes introduced by Qiskit 0.32, 15 out of 16 were verified automatically. The failed one uses an ECR gate that Giallar did not model. We added the
symbolic execution and the corresponding rewrite rule
for this new gate feature to enable its verification.
\end{reviseg}

\para{Limitations}
Compared with the manual verification framework~\cite{voqc}, 
\framework{} has a larger trusted computing base, including
the \framework{} implementation,
the Z3 SMT solver, the Coq proof checker,
the symbolic execution of Z3Py,
and the equivalence of data structures defined in Python, Coq and Z3.
We note that such a larger trusted computing base is common in prior work on push-button verification~\cite{hyperkernel,serval}.

\framework{} matches loops against a pre-defined set of loop templates.
While applicable to the current \qiskit{} compiler, 
\framework{} may fail to handle future compiler versions with 
new patterns of loops.
{\reviseg We would like to explore
how to integrate previous work on  loop invariant learning~\cite{G-CLN,ryan2019cln2inv}
to remove the loop templates in the future.}

\begin{reviseg}
\framework{}'s rewrite rule set is incomplete
and may not be able to prove the equivalence for some future
passes.
The symbolic execution for quantum circuits is also incomplete
and does not model some gate features.
New rewrite rules and more gate support may be needed
to support future \qiskit{} passes.
Besides, the
\end{reviseg}
rewrite rules for quantum circuits in \framework{} only supports local equivalence of quantum states. 
It does not support non-local 
quantum circuit optimizations that 
are implemented using other quantum state representations, such as the phase polynomial representation~\cite{Amy2017, Amy2019, nam2018automated} and the state vector quantum assertion~\cite{haner2018using}.

Besides, \framework{} is only designed to prove that
the compiled quantum circuits
preserve the semantics of the input circuits
and satisfy the topological constraints given by the coupling map.
Non-critical statements and function invocations which do not affect
the generated circuits will be discarded during the preprocessing,
such that the correctness of these statements and the generated  debugging information is not verified.

\section{Related Work}
\label{sec:related}

\para{Verified quantum compilation}
{\reviseg Following the approach of verifying classical compilation~\cite{Compcert,gu2015certified,gu2018certified},}
previous efforts on verifying quantum compilation mainly 
rely on interactive theorem provers to construct manual proofs.
ReVerC~\cite{Amy2017} is a verified compiler for reversible circuits, a subset of quantum circuits that are easier to reason about. The verification is done in $F^*$ ~\cite{fstar}. ReQWire~\cite{Rand} verifies ancillae uncomputation, one specific step of quantum compilation, using Coq~\cite{Coq12}. VOQC~\cite{voqc} verifies several quantum optimization passes such as circuit mapping and gate cancellation using Coq. Compared with \framework{}, these frameworks have the benefit of a smaller trusted computing base, and may support more complicated compilation passes whose proof goals are not provable by an SMT solver. 
However, these frameworks require both formal verification expertise %
and a significant proof burden. Moreover, all these works only verified one static version of their passes, and their proofs cannot adapt to the fast changing real-world quantum compilers such as \qiskit{}.
In contrast,  \framework{} is a push-button verification toolkit, allowing developers without much formal verification background
to write and verify compiler passes. With \framework{}, newly developed or modified passes can be easily introduced and automatically verified.

\para{Equivalence checking for quantum circuits}
Algorithms to perform efficient equivalence checking  for quantum circuits have been discussed from the view of quantum algorithms \cite{Viamontes2007}, quantum communication protocols \cite{CarlosGarcia2011}, and verification of compilation \cite{Amy2019}. Given two concrete quantum circuits $C_1$ and $C_2$, these checking algorithms can answer if $C_1$ and $C_2$ are equivalent. However, 
to verify a quantum compiler, we must prove that for \emph{any} quantum circuit $C$, the compiled circuit $C'$ is equivalent to $C$. This is beyond reach for existing checking algorithms, but can be efficiently handled by \framework{}'s rewrite rules.

\para{Verified quantum programs} 
There are several quantum programming environments that support verifying the correctness of quantum programs such as QWire~\cite{qwire}, $Q\ket{SI}$~\cite{Liu2018} and QBricks~\cite{qbricks}, 
\begin{revise}
or verifying the error bounds such as the logic of quantum robustness~\cite{lqr} and Gleipnir~\cite{gleipnir}.
\end{revise}
These frameworks are built for verifying  quantum programs
rather than for quantum compilers that transform quantum programs but are themselves classical programs.

\para{Automatic translation validation}
In classical computing, there is also a long line of work of automatic translation validation \cite{necula2000translation}, which verifies the semantic equivalence of given source and target programs in a specific translation. Automatic translation validation can be used to provide guarantee to a complicated compiler that is hard to directly verify, even when complex loops are involved \cite{dahiya2017black, gupta2020counterexample}.
However, it cannot guarantee the correctness of the compiler for \emph{all} source programs and has to be run at each compilation.
In contrast, \framework{} verifies the correctness of the \qiskit{}
compiler and the  verification was done once and for all.

\para{Push-button verification in classical computing} 
Push-button verification has been applied to building  verified compilers,  file systems,  OS kernels, and system monitors~\cite{alive,yggdrasil,hyperkernel, serval}. Many of our technical designs and verification ideas are heavily influenced by these previous works. However, these verification frameworks 
cannot support unbounded loops, nor are they directly applicable to quantum circuits.

\section{Conclusion}
\label{sec:conclusion}
\framework{} is a push-button verification toolkit that, for the first time,
enables automated verification for real-world quantum compiler passes. 
To verify unbounded loops, \framework{} introduces loop templates whose loop invariants can be automatically generated. 
\framework{} introduces the quantum circuit rewrite rules to efficiently check
the equivalence of quantum circuits,
and provides a verified library for writing new compiler passes and modularizing their proofs.
Using \framework{}, we have successfully verified \numPassVerified{} (out of 56) passes of  \qiskit{},
 the most widely used Quantum compiler,
 and detected three crucial bugs (two are quantum-specific) in the original \qiskit{} implementation.
The approach we establish with \framework{} paves the way for end-to-end verification of a complete quantum software toolchain, an important step towards practical near-term quantum computing.

\section*{Acknowledgments}
We thank our shepherd, Sorav Bansal, and the anonymous reviewers for valuable feedbacks  that help improving this
paper. We  thank Bryce Monier and John Zhuang Hui for conducting parts of the experiment
and providing helpful comments on earlier drafts. 
This work is funded in part by three Amazon Research
Awards, DARPA contract N66001-21-C-4018,  and
NSF grants CCF-1918400, CNS-2052947, and
CCF-2124080; in part
by EPiQC, an NSF Expedition
in Computing, under grants CCF-1730082/1730449; in part
by STAQ under grant NSF Phy-1818914; in part by NSF
Grant No. 2110860; by the US Department of Energy Office 
of Advanced Scientific Computing Research, Accelerated 
Research for Quantum Computing Program; and in part by 
NSF OMA-2016136 and the Q-NEXT DOE NQI Center.
Ronghui Gu is the Founder of and has an
equity interest in CertiK.
Frederic T. Chong is Chief Scientist at Super.tech and an
advisor to Quantum Circuits, Inc.

\bibliographystyle{ACM-Reference-Format}
\bibliography{references}

\end{document}